\documentclass[lettersize,journal]{IEEEtran}
\usepackage{amsmath,amsfonts}
\usepackage{algorithmic}
\usepackage{algorithm}
\usepackage{array}
\usepackage[caption=false,font=normalsize,labelfont=sf,textfont=sf]{subfig}
\usepackage{textcomp}
\usepackage{stfloats}
\usepackage{url}
\usepackage{verbatim}
\usepackage{graphicx}
\usepackage{cite}
\usepackage{url}
\usepackage{multirow}
\usepackage{xcolor}

\begin{document}
\title{Wireless-Friendly Window Position Optimization for RIS-Aided Outdoor-to-Indoor Networks based on Multi-Modal Large Language Models}

\author{Jinbo~Hou,
        Kehai~Qiu,~\IEEEmembership{Member, IEEE,}
        Zitian~Zhang,
        Yong~Yu,
        Kezhi~Wang,~\IEEEmembership{Senior Member, IEEE,}
        Stefano~Capolongo,
        Jiliang~Zhang,~\IEEEmembership{Senior Member, IEEE,}
        Zeyang~Li,
        Jie~Zhang,~\IEEEmembership{Senior Member, IEEE}
\thanks{Jinbo Hou and Jie Zhang are with the Department
of Electronic and Electrical Engineering, the University of Sheffield, Sheffield, UK (e-mail: jhou9@sheffield.ac.uk, jie.zhang@sheffield.ac.uk).
Kehai Qiu is with the Department of Computer Science and Technology, University of Cambridge, Cambridge CB3 0FD, U.K. (e-mail: kq218@cam.ac.uk), Kehai Qiu is also with Brunel University London.
Zitian~Zhang is with the School of Information and Electronic Engineering, Zhejiang Gongshang University, Hangzhou, China (e-mail: zitian.zhang@mail.zjgsu.edu.cn).
Yong Yu and Stefano Capolongo are with the Design \& Health Lab, Department of Architecture, Built Environment and Construction Engineering, Politecnico di Milano, Italy (email: yong.yu@polimi.it, email: stefano.capolongo@polimi.it). 
Kezhi Wang is with the Department of Computer Science, Brunel University London, Uxbridge, Middlesex, UB8 3PH (email: kezhi.wang@brunel.ac.uk).
Jiliang Zhang is with The State Key Laboratory of Synthetical Automation for Process Industries and The College of Information Science and Engineering, Northeastern University, Shenyang, China. (e-mail: zhangjiliang1@mail.neu.edu.cn).
Zeyang Li is with the Centre for Wireless Innovation, Queen’s University of Belfast, Belfast, BT3 9DT, UK (e-mail: lizeyang321@gmail.com).}

\thanks{Kehai Qiu, Kezhi Wang and Jie Zhang would like to acknowledge the partial support from the Eureka COMET / Innovate UK project (No: 10099265). Zeyang Li and Jie Zhang are the corresponding co-authors. Kezhi would like to acknowledge the support in part by the Royal Society Industry Fellowship (IF$\setminus$R2$\setminus$23200104)}}

\markboth{submitted to IEEE jounals for possible publication}%
{Shell \MakeLowercase{\textit{et al.}}: Bare Demo of IEEEtran.cls for IEEE Journals}

\maketitle

\begin{abstract}  
This paper aims to simultaneously optimize indoor wireless and daylight performance by adjusting the positions of windows and the beam directions of window-deployed reconfigurable intelligent surfaces (RISs) for RIS-aided outdoor-to-indoor (O2I) networks utilizing large language models (LLMs) as optimizers. 
Firstly, we illustrate the wireless and daylight system models of RIS-aided O2I networks and formulate a joint optimization problem to enhance both wireless traffic sum rate and daylight illumination performance. 
Then, we present a multi-modal LLM-based window optimization (LMWO) framework, accompanied by a prompt construction to optimize the overall performance in a zero-shot fashion, functioning as both an architect and a wireless network planner.
Thirdly, we propose an LLM-empower heuristic (LHS) framework to recommend suitable heuristic optimization algorithms and empower them through enhanced initialization and iterative refinement.  
Finally, we analyze the optimization performance of the LMWO and LHS frameworks and the impact of the number of windows, room size, number of RIS units, and daylight factor.
Numerical results demonstrate that our proposed LMWO and LHS frameworks can achieve outstanding optimization performance in terms of initial performance, convergence speed, final outcomes, and time complexity, compared with classic AI and standalone heuristic optimization methods.
The building's wireless performance can be significantly enhanced while ensuring indoor daylight performance.

\end{abstract} 


\section{Introduction}
\IEEEPARstart{I}{n} 5G wireless communication systems, approximately  90\% traffic occurs within buildings, posing unprecedented challenges to the design of indoor wireless networks.
Moreover, building wireless performance (BWP) is greatly impacted and constrained by building characteristics such as windows, layouts, and materials as demonstrated in \cite{paper_zhang2022BWP1}. 
Therefore, the wireless performance of a building should be considered and optimized according to the demand of indoor communication networks during the building's design phase. 
This is particularly crucial for reconfigurable intelligent surfaces (RIS)-aided outdoor-to-indoor (O2I) communication networks, where indoor traffic demands are met by outdoor base stations (BSs) through intermediate transmissive reconfigurable intelligent surfaces (T-RISs)\cite{paper_DAi2017_metasurfaceRIS原理}.
In such networks, walls and windows significantly impact the deployment strategy of T-RISs when mitigating the penetration loss and shadowing effects of buildings \cite{paper_Zhang2022_穿墙loss}.

Currently, most researchers proposed deploying T-RISs on the walls of buildings \cite{paper_Nemati2021_RIS_onthewall_1}.
The authors in \cite{Liang2022_RIS_onthewall_4} designed a dual-hop hybrid RIS-aided mmWave system featuring one passive and one active RIS on the surface of a building to maximize the signal-to-noise ratio of users by jointly optimizing the reflecting coefficients of the RISs and the beamforming vectors of the BSs. This optimization problem was divided into three sub-problems and solved using alternating optimization methods.
The research in \cite{Aung2023_RIS_onthewall_8} proposed a policy gradient reinforcement learning method to jointly control the beamforming power of each user and the phase shifts of RISs to maximize spectral efficiency for both indoor and outdoor users in a wall-deployed simultaneously transmitting and reflecting (STAR)-RIS-assisted downlink communication system.
Similarly, researchers in \cite{paper_He2023_RIS_onthewall_2} addressed a wall-deployed STAR-RIS-aided three-dimensional indoor and outdoor localization problem by optimizing power allocation between refraction and reflection, as well as between two mobile stations, using a principal angle analysis method.
Additionally, the research \cite{Li2022_RIS_onthewall_10} investigated a wall-deployed hybrid double-RIS-aided and relay-aided system to eliminate the impact of building occlusion and penetration loss in O2I communication networks by proposing two closed-form algorithms for passive RIS beamforming.

Unfortunately, these wall-deployed T-RISs still experience significant penetration loss caused by concrete walls \cite{Concrete_research}. Moreover, wall-deployed RISs are difficult to reconfigure, replace, or repair due to high structural modification costs and the potential hazards associated with the rapid evolution of RIS-related technologies\cite{news_承重墙损坏}.
Consequently, windows, as the most flexible components of a building, are promising to play a crucial role in RIS-aided O2I communication networks, as highlighted by recent research.
Researchers in \cite{Kitayama2023_RIS_onthewindwo_3} proposed a window implementation strategy using a transparent metasurface lens, facilitating installations on glass windows and other visually sensitive areas. This approach allowed the scattering characteristics of metasurfaces to be dynamically controlled, mitigating coverage holes in an O2I network while maintaining high optical transparency.
Subsequently, a window-deployed active intelligent transmitting surface solution was proposed in \cite{XX2023_RIS_onthewindow_5} in an O2I network to optimize the indoor weighted traffic sum-rate. 
Additionally, window-deployed transparent active RISs were proposed in \cite{tranparent—RIS} for uplink enhancement in indoor-to-outdoor mmWave communication networks using closed-form optimization methods.
These RISs can be integrated into optically transparent glass windows, providing environmental and aesthetic benefits without compromising visual effects.

\begin{figure}[!t]
    \centering
    \includegraphics[width=3.2 in]{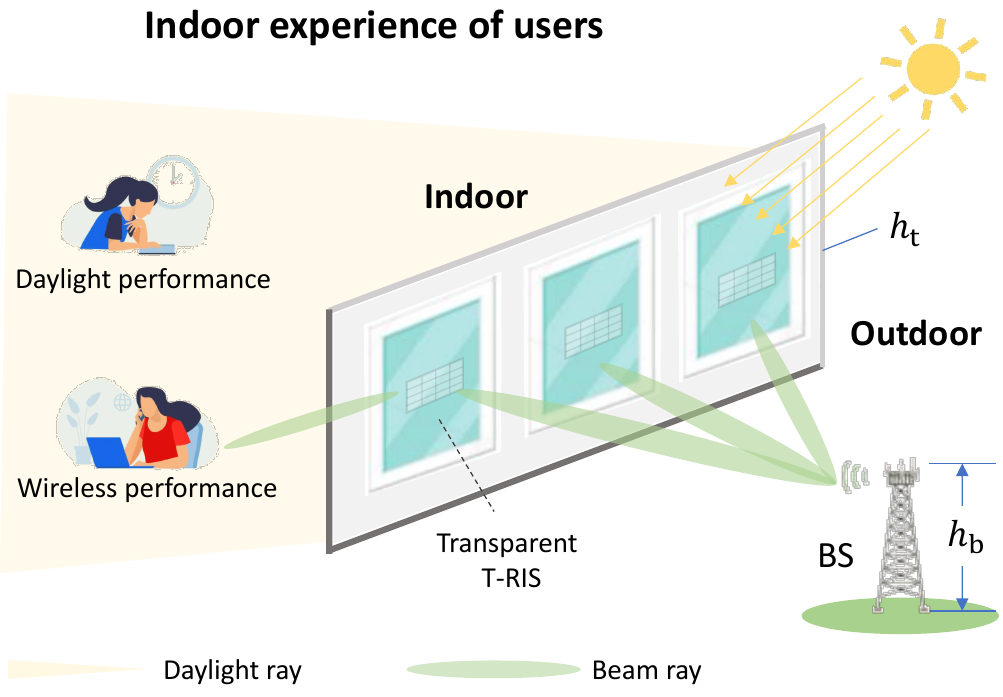}
    \caption{Illustration of the indoor user's well-being experience considering the performance of wireless and daylight via transparent T-RISs.}
    \label{ideal}
\end{figure}

However, these studies have primarily focused on the deployment of RISs while neglecting the impact of window attributes, particularly the positioning of windows.
Inadequate window placement can result in the suboptimal placement of T-RISs, leading to a deterioration in indoor wireless communication performance, which subsequently affects the quality of experience (QoE) of indoor users. 
Meanwhile, the window placement is also crucial for indoor daylight performance \cite{paper_Aesthetic_window}, which significantly influences indoor users' QoE in terms of their mood, morale, mental state, and health \cite{light_on_human}.

Therefore, in this paper, we first investigate a passive T-RIS-aided O2I network with indoor blockages and aim to optimize both indoor wireless and daylight performance simultaneously by adjusting the positions of windows to enhance the overall QoE of indoor users as shown in Fig. \ref{ideal}.  
We then formulate a joint optimization problem that addresses indoor wireless performance and daylight performance, subject to constraints on the minimum interval of windows and minimum indoor daylight requirements.
To tackle this problem, we design large language model (LLM)-based window optimization (LMWO) and LLM-power heuristic (LHS) frameworks, followed by specific simulations and discussions. Our contributions can be summarized as follows:

$\bullet$ To our best knowledge, this is the first work to concurrently optimize the performance of the wireless network and daylight system in a passive T-RIS-aided O2I network with a specific focus on the role of windows in enhancing the well-being QoE of indoor users through a problem formulation. 

$\bullet$ Secondly, we propose a novel LLM-based optimization framework, the LMWO framework, to address this challenge by leveraging the pre-training knowledge, understanding, and reasoning capabilities of LLMs.
Within this framework, an initialization module and an optimization module are designed to iteratively improve the performance of the LLM-based optimizer by interacting with the target environment after incorporating redundant information and environmental feedback as a fusion of multi-modal prompts (e.g., text and image).

$\bullet$ Thirdly, we propose the LHS framework to enhance the optimization performance by integrating heuristic optimization algorithms with LLMs. In this framework, LLMs recommend novel heuristic algorithms, propose initial solutions, and iteratively refine solutions for target problems by exploiting the optimization knowledge and programming capabilities of LLMs. 
By incorporating heuristic search strategies, LLMs are equipped with well-defined search procedures, which balance the trade-off between optimization efficiency and solution stability.

$\bullet$ Finally, simulation results demonstrate that our proposed LMWO and LHS frameworks significantly outperform the LLM-proposed heuristic algorithm and reinforcement learning (RL) optimization method in terms of initial performance, convergence speed, time complexity, and final performance.
Additionally, we investigate the impact of the number of windows, the number of RIS units, room size, and daylight factor on the performance of the LMWO framework.

The rest of this paper is structured as follows: Section II discusses the system models for both the wireless system and daylight system with a joint optimization problem formulation. 
Section III and IV detail the basic knowledge of LLMs and our proposed LMWO and LHS frameworks. 
Section V includes an experimental setting and a discussion of the results with an analysis of influence elements. 
Finally, we conclude our work and present potential future directions in Section VI.

\section{System Model and Problem Formulation} 
In this section, we introduce the wireless network model of a passive T-RIS-aided O2I network, followed by a daylight evaluation system for the indoor scenario. 
Combining these two systems, a joint optimization problem is formulated considering both wireless traffic sum rate and daylight illumination performance.  


\subsection{Wireless Network Model}
\begin{figure}[!t]
    \centering
    \includegraphics[width=2.9in]{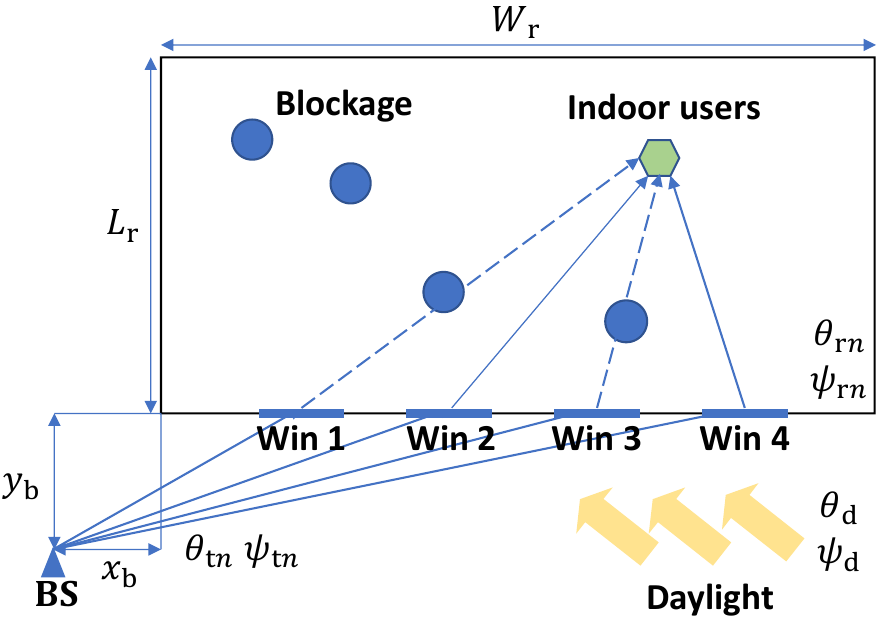}
    \caption{The wireless network and daylight system from a top-down perspective}, where $\theta$ and $\psi$ represent the elevation and azimuth angle of corresponding links, respectively, and "Win" is an abbreviation for “window”.
    \label{system model}
\end{figure}

3-dimensional (3D) wireless network model of a typical passive T-RIS-aided O2I network is built on our previous research\cite{paper_li2022coverage} as shown in Fig. \ref{system model}. 
The focused rectangular workplace room has width $W_{\rm r}$, length $L_{\rm r}$, altitude height $H_{\rm r}$, and $N$ windows on its long side wall of the near end. All windows have a consistent size of $L$ length and $W$ width.
These windows are fixed and remain closed, which is common in commercial centres and office buildings to ensure safety, air quality, and noise control.
In particular, $U$ passive T-RIS units are uniformly deployed at the centre of windows with a fixed beam orientation to enhance the wireless communication between an outdoor BS with $N_{\rm t}$ antennas and an indoor user with a single antenna. 
We assume that the transmission from the BS to each RIS of windows is non-blocked and the direct transmission between the BS and UE is not considered due to the huge penetration loss. 
Specifically, the BS is $H_{\rm t}$ altitude height and located at (0,0) in a 2-dimension Cartesian coordinate system. The left near point of the building is located at $(x_{\rm b}, y_{\rm b})$. 
The central positions of windows are presented as $\mathcal{P}=\{(x_{1}, y_{\rm b}),(x_{2}, y_{\rm b}),..., (x_{N},y_{\rm b})\}$.
The beam elevation angles and azimuth angle of window-deployed RISs are denoted as ${\Theta}_{\rm r} = \{\theta_{{\rm r}1},\theta_{{\rm r}2},...,\theta_{{\rm r}N}\}$ and ${\Psi}_{\rm r} = \{\psi_{{\rm r}1},\psi_{{\rm r}2},...,\psi_{{\rm r}N}\} $, respectively.

Inside the room are $M$ uniform measurement points to assess indoor wireless performance. 
For each measurement point, there is a user weight value to present the probability of the user appearing at this point following the function of the 2-dimensional joint beta distribution, which can represent the probability of traffic demand requirements and is presented as $\mathcal{W} = \{w_{\rm 1},w_{\rm 2},..., w_{M}\}$. 
Meanwhile, indoor environmental elements like furniture, pillars, and decorations are considered to be blockages, which are modelled as cylinders with an average radius of $R$. The distribution of the blockages' centres follows the Poisson point process (PPP) with a density $\rho$.
Considering the high penetration loss of indoor blockages, only line-of-sight (LoS) transmission is considered \cite{paper_Zhang2022_穿墙loss,block_ZYX}. 
Thus, according to the propagation model in \cite{New_model}, the average received power of the signal transmitted by the T-RIS on the $n$th window is shown as Eq. (\ref{AR_Power}).

\begin{equation}
    \label{AR_Power}
    \centering
    \small
    P_n = \frac{P_{\rm t}N_{\rm t}^{2}d_{\rm x}^{2}d_{\rm y}^{2}\lambda^{2}\cos\theta_{{\rm t}n}\omega^2(\sum_{i=1}^{\sqrt{U}}\sum_{j=1}^{\sqrt{U}}{\rm exp}(-j\phi_{n,i,j}))^2}{16\pi^2d_{n1}^2d_{n2}^2},
\end{equation}

\noindent where $P_{\rm t}$ is the transmit power of the signal from the BS, $d_{\rm x}\times d_{\rm y}$ is the size of the units of passive T-RISs on the $n$th window, $\lambda$ is the wavelength of the carrier signal,  $\omega$ is the penetration loss when the signal passes through the window, $d_{n1}$ is the distance between the $n$th window and the BS, $d_{n2}$ is the distance between the $n$th window and the user, respectively,
$\phi_{n,i,j}$ is the phase of the wave reflected by the $i$th row and $j$th column unite on the RIS of the $n$th window and can be calculated as:  

\begin{equation}
    \label{phase_c}
    \centering
    \phi_{n,i,j} = \phi_{{\rm a},n,i,j} - \phi_{{\rm r},n,i,j},
\end{equation}

\noindent where $\phi_{{\rm a},n,i,j}$ is the phase shift of the wave due to the location of the $i$th row and $j$th column unite on the RIS of the $n$th window as shown in Eq. (3), $\phi_{{\rm r},n,i,j}$ is the field pattern of the $i$th row and $j$th column unit on the RIS of the $n$th window as shown in Eq. (4). $\theta_{{\rm t}n}$ and $\psi_{{\rm t}n}$ are the elevation and azimuth angle from the BS antennas to the centre of the RIS of the $n$th window, $\theta_{{\rm r}n}$ and $\psi_{{\rm r}n}$ are the elevation and azimuth angle from the centre of the RIS of the $n$th window to the UE, $\theta_{{\rm r}n}^{'}$ and $\psi_{{\rm r}n}^{'}$ are the desired elevation and azimuth angle from the centre of the RIS of the $n$th window to the UE.

\begin{figure*}[hb]
    \centering
    \label{equ:phi1}
    \begin{equation}
        \phi_{{\rm a},n,i,j}= {\rm mod}(-\frac{2\pi}{\lambda}(\sin{\theta_{{\rm t}n}}\cos{\psi_{{\rm t}n}}+\sin{\theta_{{\rm r}n}}\cos{\psi_{{\rm r}n}})(m-\frac{1}{2})d_{\rm x}+(\sin{\theta_{{\rm t}n}}\sin{\psi_{{\rm t}n}}+\sin{\theta_{{\rm r}n}}\sin{\psi_{{\rm r}n}})(n-\frac{1}{2})d_{\rm y}),2\pi).
    \end{equation}    
\end{figure*}

\begin{figure*}[hb]
    \centering
    \label{equ:phi2}
    \begin{equation}
        \quad \  \phi_{{\rm r},n,i,j}= {\rm mod}(-\frac{2\pi}{\lambda}(\sin{\theta_{{\rm t}n}}\cos{\psi_{{\rm t}n}}+\sin{\theta_{{\rm r}n}^{'}}\cos{\psi_{{\rm r}n}^{'}})(m-\frac{1}{2})d_{\rm x}+(\sin{\theta_{{\rm t}n}}\sin{\psi_{{\rm t}n}}+\sin{\theta_{{\rm r}n}^{'}}\sin{\psi_{{\rm r}n}^{'}})(n-\frac{1}{2})d_{\rm y}),2\pi). \quad     
    \end{equation}
\end{figure*}

Subsequently, the received traffic data rate of the measurement point $m$ is derived by:

    \begin{equation}
    \label{eq:sum_rate}
    \begin{aligned}
    \small
    \gamma_{m} = \sum_{z\in{\mathcal{Z}}}P_{\rm LoS}(z)B{\rm \log}\left(1+\frac{(\sum_{n\in{z}} \sqrt{P_n})^2}{\sigma_{m}^{2}}\right),
    \end{aligned}
    \end{equation}

\noindent where $z$ is the set of unblocked RISs, $\mathcal{Z}$ is the possible combinations of RISs, e.g., $\{(1,2,3), (1,2), (1,3),(1), (2), (3)\}$ when there are three window-deployed RISs, $P_{\rm LoS}(z)$ is the probability that the RISs from the set $z$ is unblocked as the derivation Eq. (14) and Eq. (20) in \cite{paper_li2022coverage}, $B$ is the channel bandwidth, $\sigma_{m}^{2}$ is the power of the additive white Gaussian noise (AWGN) at measurement point $m$.

\subsection{Daylight Evaluation Model}
For the indoor daylight analysis, daylight performance is evaluated using ray-tracking algorithms implemented through the Radiance and Daysim plugins within the Rhinoceros 3D Grasshopper software environment \cite{grossper_theory}.
The ray-tracing method offers the most accurate and detailed results \cite{light_on_human}, considering the diverse daylight sources, the diffuse reflection phenomena, and the complexity of light propagation paths. 
Specifically, Radiance is a lighting simulation tool based on hybrid deterministic-stochastic ray tracing approaches \cite{radiance_theory} developed by Greg Ward Larson at Lawrence Berkeley National Laboratory.
Radiance provides a realistic representation of light interactions by incorporating weather and illumination data into the indoor simulation model in compliance with EU Building Design standards \cite{indoor_light_design}.
Subsequently, Daysim analyses daylight performance using Radiance algorithms. 
Additionally, natural weather and illumination data are introduced from the Ladybug \cite{lady_bug}, an open-source project focused on environmental design.


Inside the room, $M$ measurement points are established to assess daylight, consistent with the wireless network model. 
For each measurement point, a single-day grid analysis is conducted on the winter solstice (the 21st of Dec., the day with the weakest daylight of a year) at 15:00, considering only external natural and sunlight. 
The elevation and azimuth angles from the sunlight source to the centres of windows are denoted by $\theta_{\rm d}$ and $\psi_{\rm d}$, respectively.
The daylight transmission rate of the windows used is presented as $\beta$ using double-layer soda-lime glass.
In terms of light reflection, the simulation calculates two bounces off the wall, floor, and ceiling surfaces with reflection coefficients, represented by $F_{\rm w}$, $F_{\rm f}$, and $F_{\rm c}$, respectively.
The illumination degree $I_{m}$ at the $m$th measurement point (lux as unit) is calculated as:

\begin{equation}
    \label{daylight_illumination}
    \centering
    \small
    I_{m} = I_{m}^{\rm l}+ \widetilde{I_{m}^{\rm l}} + I_{m}^{\rm s} + \widetilde{I_{m}^{\rm s}},
\end{equation}

\noindent where $I_{m}^{\rm l}$ and $\widetilde{I_{m}^{\rm l}}$ are the direct and indirect illumination from the skylight, respectively, $I_{m}^{\rm s}$ and $\widetilde{I_{m}^{\rm s}}$ are the direct and indirect illumination from solar radiation, respectively. In particular, the direct ones include the direct light and solar illumination transmitted through windows. The indirect ones include the inter-reflected light, both internal and external reflections.
Consequently, the output set of daylight illumination levels for all measurement points is presented as $\mathcal{I}=\{I_{1}, I_{2},..., I_{M}\}$.

\subsection{Problem Formulation}

To improve the overall QoE of indoor users, wireless performance and daylight performance are simultaneously optimized by adjusting the window positions and beam directions of window-deployed RISs according to an indoor user weight matrix $\mathcal{W}$.
The optimization results are compared against those of the common uniform-distributed window (UDW) strategy, where windows are uniformly implemented with beam directions perpendicular to the surface. The improvement of wireless performance $\phi_{\rm w}$ is calculated as:

\begin{equation}
    \label{Sum_rate}
    \centering
    \phi_{\rm w} =  \frac{\sum_{m=0}^{M} \gamma_{m} / \gamma_{m}^{'} \cdot w_{m}}{M},      
\end{equation}

\noindent where $\gamma_{m}^{'}$ presents the received traffic data rate at the $m$th measurement point in UDW strategy.
Similarly, the improvement of daylight performance $\phi_{\rm d}$ is calculated as: 
\begin{equation}
    \label{d_impro}
    \centering
    \phi_{\rm d} = \frac{\sum_{m=0}^{M} I_{m} / I_{m}^{'} \cdot w_{m}}{M},
\end{equation}
\noindent where $I_{m}^{'}$ presents the daylight illumination degree at the $m$th measurement point in UDW strategy.
Finally, the optimization problem is formulated as:

\begin{equation}
    \label{w_impro}
    \begin{aligned}
    \phi_{\rm o} = & \max\limits_{\mathcal{P}, {\Theta}_{\rm r}, {\Psi}_{\rm r}}\ (\phi_{\rm w} + \eta \cdot \phi_{\rm d}),\\
    & \ \ \ \ {\rm s.t.} \ \ \phi_{\rm d} \geq \mathcal{T_{\rm min}}, \\\
    &\ \ \ \ \quad \quad x_{n+1} - x_{n} \geq d_{\rm min}, \forall x_{n}\in \mathcal{P},
    \end{aligned}
\end{equation}

\noindent where $\phi_{\rm o}$ is the overall QoE optimization performance, $\eta$ is adopted as daylight factor to control the optimization tendency, where higher values indicate a greater focus on sunlight, $\mathcal{T_{\rm min}}$ is the minimum daylight improvement requirement to guarantee indoor daylight performance,  $d_{\rm min}$ presents the minimum distance between centres of windows, which is larger than the window width for aesthetic considerations.

This optimization problem is a black-box and non-convex problem caused by the following factors:
i. The complexity and non-convexity caused by the blockage distribution and user distribution;
ii. the sophisticated nature of the realistic ray tracing model employed for the daylight evaluation system; and iii. The constraints imposed by variables and systems. Existing heuristic algorithms, surrogate-based algorithms, and reinforcement learning (RL) methods suffer from being trapped in local suboptimal solutions, insufficient search efficiency, and huge training costs. Moreover, these optimization algorithms require manual customization tailored to specific environments, thereby restricting their generalization capability and automation.

To address these challenges, we propose two LLM-based optimization frameworks, leveraging LLMs' pre-trained knowledge, reasoning ability, and continuous learning capacity in black-box optimization problems.   
Unlike conventional optimization approaches that require step-by-step programming, LLM-based optimizers do not require detailed and precise optimization execution steps when interacting with environments and receiving feedback.
Moreover, LLM-based optimizers offer potentially effective solutions for complex optimization problems thanks to their extensive pre-trained data, unique multi-modal understanding mechanisms, and continuous learning capabilities.
According to \cite{LLM_algorithm_2, LLM总体分析}, we summarize four prime advantages of LLM-based optimizers, compared with existing optimization methods:
i. zero-shot capability \cite{Survey_AI_RIS} and warm start performance;
ii. improved final performance;
iii. reduced time and faster convergence; 
and iv. model-free nature and generalization ability,
which align well with the requirements of our focused optimization problem.

%
\section{Methodology}
\begin{figure*}
    \centering
    \includegraphics[width=5.1in]{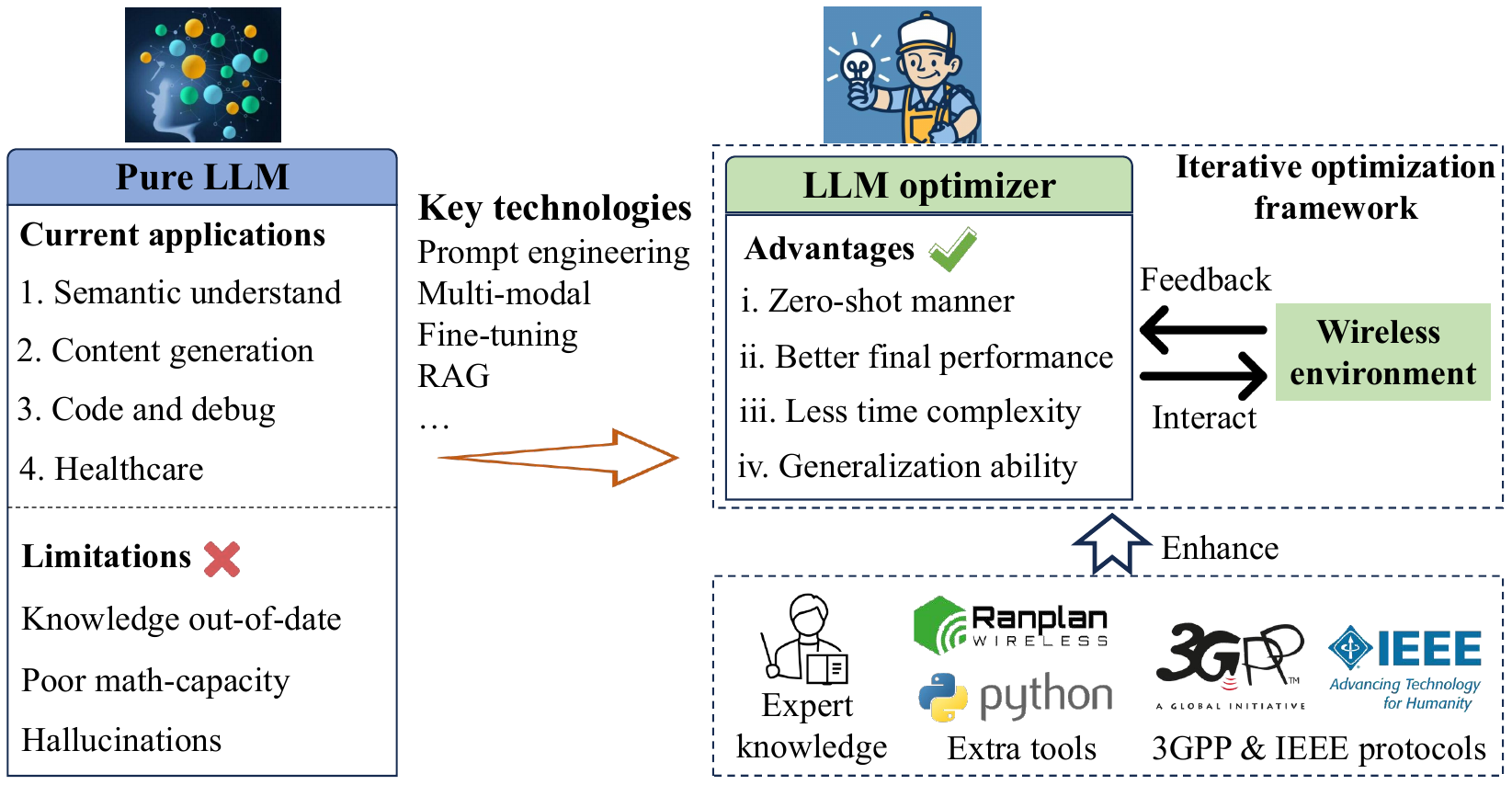}
    \caption{The overview of LLM-based optimizers.}
    \label{LLM_idea}
\end{figure*}

In this section, we first introduce the foundation of LLMs as optimizers as depicted in Fig. \ref{LLM_idea}, accompanied by a concise literature review of their current applications.
Then, we present the design of the LLM prompt template and multi-modal feedback for our focused optimization problem, followed by a specific demonstration of the LMWO framework.

\subsection{LLM as Optimizer}

Recently, LLMs have garnered tremendous attention and are famous for ChatGPT, enabled by transformer technology \cite{LLM_survey}.
LLMs are distinguished by their ability to generate human-like contextually appropriate and grammatically correct outputs due to their extensive parameters, trained on vast text datasets covering diverse domains of human life \cite{GPT训练}.
Consequently, it is plausible that LLMs contain human-like capacities of continuous learning through trial and error and exhibit reasonable decision-making abilities, making them potential ideal optimizers for various optimization problems across different domains.

As demonstrated in \cite{LLM_optimizer_s2}, there are two primary approaches to utilizing LLMs as optimizers: 
In black-box optimization problems, LLMs can serve as solution generators, producing reasonable solutions immediately according to the information and instructions provided through natural language prompts \cite{LLM_optimizer_s1}. Moreover, LLMs can collaborate with widely used heuristic optimization algorithms by suggesting suitable solutions and evolutionary directions, thereby enhancing efficiency in complex black-box optimization problems \cite{LLM_algorithm_2}.
For specific optimization problems, LLMs can function as optimization algorithm selectors and generators to understand the characteristics of a given problem and subsequently design appropriate optimization code, including hyperparameter tuning \cite{hyper} and algorithm steps \cite{CoT}.

However, the application of LLMs as optimizers is significantly limited by issues such as hallucination answers (e.g., incorrect and nonsensical content), knowledge out-of-date, and limited mathematical capacities \cite{LLM_limitation}.
To address these limitations, a robust framework and proper prompt design are necessary to activate the optimization capabilities of LLMs, ensuring accurate, knowledge-aligned, and condition-compliant output. 
Specifically, prompts, which are text strings designed for interaction with LLMs, allow LLMs to integrate user-defined parameters of target systems, provide essential information on optimization problems, leverage professional simulation results from external tools (e.g., Python, Matlab, and Ranplan \cite{Ranplan}), and incorporate human instructions.  

Currently, LLM-related optimization methods have shown impressive optimization capabilities across various research domains, including essay writing, code explanation, debugging, education, customer service, content creation, and healthcare \cite{LLM_algorithm, LLM_algorithm_2, LLM_algorithm_application2, LLM_algorithm_application4}. 
In the wireless communication field, LLM-related research is advancing rapidly.
The authors in \cite{LLM_wireless_magazine} proposed a comprehensive WirelessLLM framework for adapting and enhancing LLMs to address wireless communication networks' unique challenges and requirements. 
A pre-trained LLM-empowered framework was proposed in \cite{LLM_intrusion_detection} to perform fully automatic network intrusion detection with three in-context learning methods to enhance the performance of LLMs without further training.
The authors in \cite{LLM_Channel_modeling} proposed a workflow for automated network experimentation relying on the synergy of LLMs and model-driven engineering, where the experiment code was generated based on the textual description.
The research in \cite{LLM_edgeAI} presented autonomous edge AI systems to organize, adapt, and optimize themselves, leveraging the power of LLMs to meet users’ diverse requirements and generating a privacy-preserving manner code to train new models. 
In \cite{LLM_AP_deploy}, the authors introduced an LLM-based optimizer for wireless network planning and optimization, focusing on optimizing the number and placement of wireless access points.
Moreover, the research in \cite{LLM_RIS} leveraged the analytical capabilities of LLMs, combined with key data to optimize RIS-based communication systems and achieve energy-efficient performance in real-time.


\subsection{LMWO Framework}

\begin{figure*}
    \centering
    \includegraphics[width=7.3in]{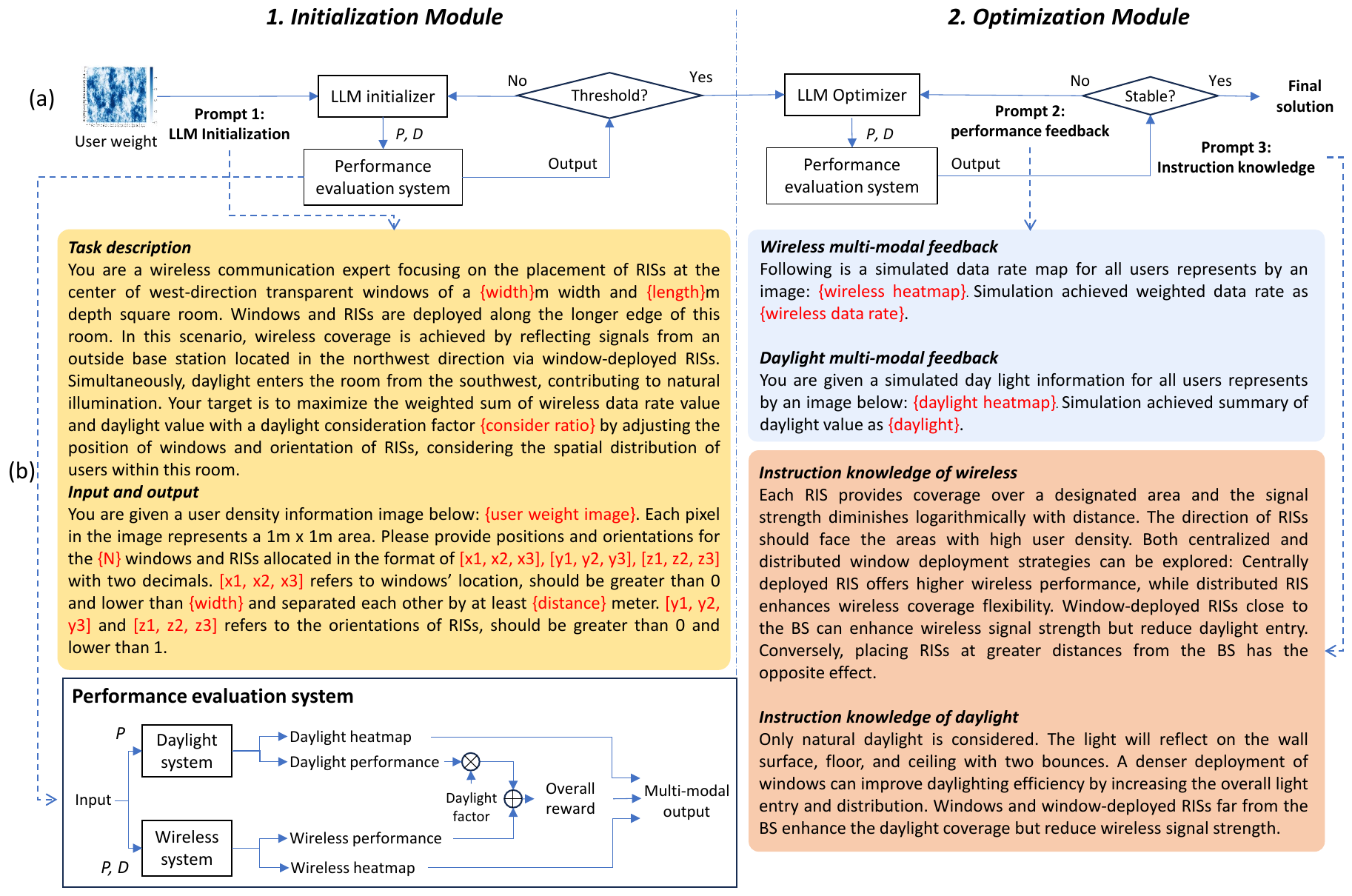}
    \caption{The structure of the LMWO framework, including two modules, three prompts, and the performance evaluation system.}
    \label{LMWO_framework}
\end{figure*}



The following introduces the workflow of the LMWO framework, including its inputs, outputs, core modules, and prompts designed for optimizing wireless and daylight performance as depicted in Fig. \ref{LMWO_framework} (a).
In this framework, prompt engineering serves as the primary approach for instructing the LLM and integrating LLM with the target environment, enabling an iterative optimization framework.

Specifically, the workflow begins with a user distribution matrix from the target room as the input. Then, an initialization module is designed to generate a qualified initialization.
Subsequently, an optimization module refines the solution iteratively until stability.   
Finally, the solution with the best optimization result is output as the final solution.
In these modules, three reproducible prompts are constructed to provide task-oriented instructions, strictly define the format, and embed environmental feedback to LLMs as instructions for the target optimization problem as shown in Fig. \ref{LMWO_framework} (b).
To supply LLMs with redundant information, we propose a prompt that includes five categories of information and organizes reproducible, structured prompts \cite{LLM_algorithm_application4}, including task description, environmental information, input \& output formats, historical experience, and instruction knowledge.
This approach ensures that comprehensive contextual information enables LLMs to better understand the optimization problem and provides sufficient cues for accurate inference.
Consequently, LLMs can generate accurate outputs while minimizing hallucination effects.


Moreover, we design the wireless environment feedback as a fusion of multi-modal information \cite{Multi-Modal_LLM}, containing wireless sum rate performance $\phi_{\rm w}$, daylight illumination performance $\phi_{\rm d}$, a heatmap figure of wireless sum rate performance at measurement points, and a heatmap figure of daylight illumination performance at measurement points as shown in Fig. \ref{LMWO_framework} (c). 
By capturing different types of environmental feedback (e.g., text and images), LLMs can establish connections between feedback data and enhance situational awareness through cross-verified information. Therefore, LLMs can provide more reasonable solutions for our optimization problems. 
Next, we discuss the details of two core modules:

\begin{algorithm}[!t]
\caption{\textbf{:} Initialization module}
\label{module 1}
\begin{algorithmic}[1]
\STATE input the \textbf{initialization prompt 1} to the LLM;
\STATE achieve $\mathcal{P}$, ${\Theta}_{\rm r}$, and ${\Psi}_{\rm r}$ from the LLM.
\STATE interact with the wireless and daylight evaluation systems to get the multi-modal output;
\IF {reward $\leq$  $\mathcal{T}_{\rm i}$}
\STATE go back to step 2;
\ENDIF
\STATE forward the initialization to the optimization module;
\end{algorithmic}
\end{algorithm}

\subsubsection{Initialization module}
The initialization module determines the window positions $\mathcal{P}$ and the RIS beam directions ${\Theta}_{\rm r}$ and ${\Psi}_{\rm r}$ through an LLM-based initializer and interactions with the wireless network and daylight evaluation system. This process involves iterations to achieve qualified results that exceed threshold $\mathcal{T}_{\rm i}$, emphasizing the significance of initial performance for the LMWO framework as illustrated in Algorithm \ref{module 1}.
Within this module: 
\begin{itemize}
\item \textbf{LLM initialization prompt} is constructed by the information of task description, environmental information, input format, and output format as shown in Fig. \ref{LMWO_framework} (b). 
This information provides details about the optimization task (e.g., variables, optimization goals, and constraints), environmental information of the target room, and sufficient instructions for the input and output with restricted format and appropriate guidance. 
\end{itemize}

\begin{algorithm}[!t]
\caption{\textbf{:} Optimization module}
\label{module 2}
\begin{algorithmic}[1]
\STATE achieve the initialization solution;
\STATE interact with the wireless system and the daylight evaluation system;

\IF {stable or maximum steps} 
\STATE end the process and output the final solution;
\ELSE{}
\STATE construct the \textbf{performance feedback prompt} by leveraging the multi-modal output and the \textbf{instruction knowledge prompt}. 
\STATE achieve $\mathcal{P}$, ${\Theta}_{\rm r}$, and ${\Psi}_{\rm r}$ from the LLM.
\STATE go back to step 2;
\ENDIF
\end{algorithmic}
\end{algorithm}

\subsubsection{Optimization module}
Subsequently, the optimization module contains a combinatorial LLM-based optimizer that iteratively updates performance based on the results from the initialization module as shown in Algorithm \ref{module 2}. 
In each iteration, 
window positions and RIS beam directions are generated and forwarded to the corresponding wireless and daylight evaluation system.
This process continues until a stable status or maximum iteration steps. 
Within the optimization module, there are two types of prompts:

\begin{itemize}
\item \textbf{Performance feedback prompt} informs the LLM about the feedback information from the wireless and daylight system model using the historical experience information as shown in Fig. \ref{LMWO_framework} (b) to iteratively adjust and improve the LLM’s response based on the feedback.
\item \textbf{Instruction knowledge prompt} uses the instruction knowledge information as shown in Fig. \ref{LMWO_framework} (b) to integrate additional indoor wireless network knowledge and indoor daylight design knowledge from physical, mathematical, and experiential aspects into the LLM and enhance its decision-making capability. Incorporating instruction knowledge equips the LLM with an enhanced capacity in wireless communication and architectural domains to address complex optimization problems.

\end{itemize}

The computational complexity of the LMWO framework is $\mathcal{O}(\alpha N_{\rm o}(N_{\rm i}^2\delta_{\rm i}+N_{\rm i}\delta_{\rm i}^2)(T_{\rm i}+T_{\rm o}))$, where $N_{\rm i}$ is the sequence length of input, $\delta_{\rm i}$ is the representation dimension of input, $\alpha$ is the layer number of the transformer, $N_{\rm o}$ is the sequence length of output, $T_{\rm i}$ and $T_{\rm o}$ are the steps of the initialization and optimization module, respectively. Specifically, the complexity of typical self-attention is $\mathcal{O}(N_{\rm i}^2\delta_{\rm i})$ and the complexity of typical feed-forward networks is $\mathcal{O}(N_{\rm i}\delta_{\rm i}^2)$ \cite{Attention}.

\subsection{LHS Framework}

\begin{figure*}
    \centering
    \includegraphics[width=7.3in]{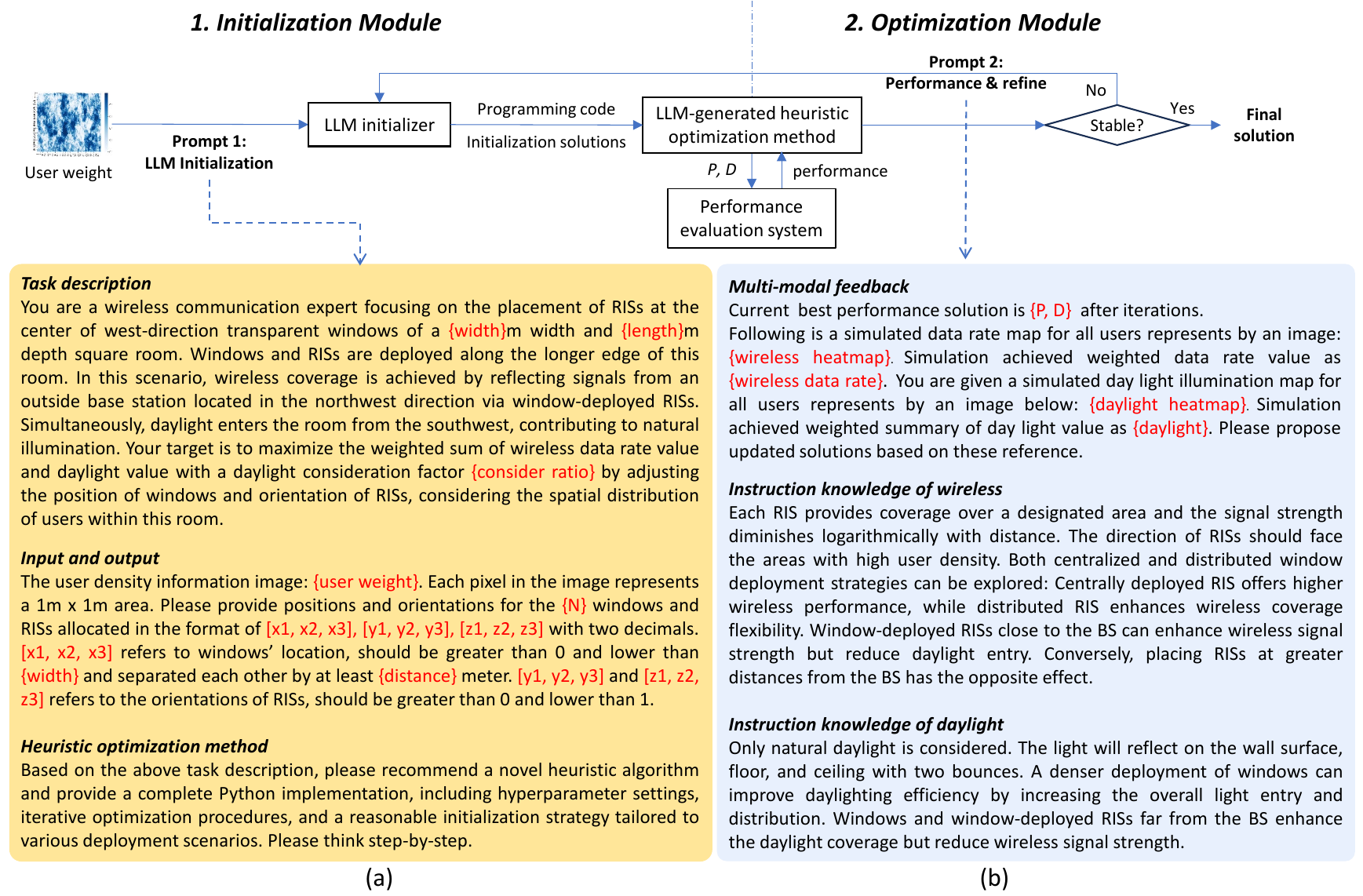}
    \caption{The structure of the LHS framework, including two modules and five prompts. The LLM initialization, performance feedback, instruction knowledge, and performance evaluation system remain the same as the ones in the LMWO framework.}
    \label{LHS_framework}
\end{figure*}

As introduced above, the heuristic algorithms are insufficient for addressing the target black-box optimization problem due to their low search efficiency, heavy reliance on domain-specific design, and limited quality of initial solutions.
On the other hand, LLM optimizers lack a well-defined and systematic search mechanism, often terminating exploration when encountering a high-performing sub-optimal solution. 
To address these limitations, this section introduces the LHS framework, which combines the strengths of heuristic and LLM optimizers. The LHS framework is designed to improve the quality of the initial population, enhance search efficiency, and increase generalization capability for solving the target black-box optimization problem.

Specifically, the LLM optimizer plays a crucial role in population initialization, optimization algorithm generation, and the reflective updates process, leveraging LLMs' reasoning capacity. 
First, the LLM optimizer recommends an appropriate heuristic algorithm and generates a logically initial population after understanding the optimization problem,    instead of traditional random or human-designed initializations. 
Subsequently, the LLM optimizer generates executable code for the selected heuristic algorithm, which can be implemented in Python, MATLAB, or other programming software.
After a predefined number of iterations or when the heuristic algorithm reaches convergence, the population is updated by the LLM optimizer by considering the results of the heuristic optimization algorithms to speed up the searching process and accelerate convergence.
Throughout these stages, the LHS framework significantly achieves searching efficiency and adaptability across various optimization problems without requiring manual intervention, benefiting from the pre-trained knowledge of LLM optimizers and the heuristic exploration strategies.
As shown in Fig. \ref{LHS_framework}, there are two core modules in the LHS framework:

\textit{1) Initialization Module:}
The initialization module is designed to propose suitable heuristic optimization methods, operation procedures, hyper-parameters, and an initial population after defining the task content, objectives, goals, and constraints through the LLM initialization prompt.  
In contrast to the initialization prompt in the LMWO framework, the prompt further guides the LLM initializer to recommend suitable novel heuristic algorithms and automatically generate corresponding implementation code, leveraging step-by-step reasoning and programming capacity of LLMs.
The initialization solutions of the proposed heuristic algorithm are also suggested by the LLM initializer and subsequently evaluated for their fitness by a performance evaluation system, as illustrated in Fig. \ref{LHS_framework}(a).

\textit{2) Optimization Module:}
The optimization module begins by executing the proposed heuristic algorithm code on simulation and modeling platforms. After a predefined number of iterations, the best-performing solution, along with its multi-modal performance in both wireless and daylight dimensions, is fed back to the LLM optimizer for updating the current population and enhancing search efficiency via a performance-driven prompt, as illustrated in Fig. 5(b). Finally, the best-performing solution will be output until the maximum LLM update steps $T_{\rm u}$ or convergence is reached.

The computation complexity of the LHS framework is $\mathcal{O}(\alpha N_{\rm o}(N_{\rm i}^2\delta_{\rm i}+N_{\rm i}\delta_{\rm i}^2))T_{\rm i} + \mathcal{O_{\rm h}}T_{\rm o}T_{\rm i}$, where $\alpha$ is the layer number of the transformer, $N_{\rm o}$ is the sequence length of output, $N_{\rm i}$ is the sequence length of input, $\delta_{\rm i}$ is the representation dimension of input, $T_{\rm i}$ and $T_{\rm o}$ are the steps of the initialization and optimization module, respectively, $\mathcal{O_{\rm h}}$ is the computation complexity of the proposed heuristic optimization algorithm \cite{Attention}.

\section{Simulation Result and Discussion}

In this section, we first introduce the simulation settings
employed in this paper. 
Then, we present the experimental results of the LMWO and LHS frameworks compared with two benchmark methods.
Finally, we discuss the influence of the number of windows, daylight factor, number of RIS units, and room size on the performance of two frameworks, followed by a time complexity analysis.

\subsection{Simulation Setting}
In this simulation, 
the room is assumed to be located in Sheffield the UK, with west-facing windows. The meteorological data of Sheffield is sourced from EnergyPlus Weather (EPW) files downloaded from the website \cite{lady_bug}. 
Moreover, we consider a rectangular room with a 10m width and 20m length. 
The detailed parameters are shown in Table \ref{table:paremeters}. 
The maximum operation steps of all comparison methods are set to 1500, which is sufficient to achieve stable convergence. Towards the LMWO and LHS frameworks, if the result does not change in five steps or reaches a maximum of ten steps, the optimization process will stop.
Additionally, each method is conducted ten times to obtain an average result to mitigate individual randomness. In each scenario, comparison methods include:

\begin{itemize}
\item \textbf{Deep deterministic policy gradient (DDPG):} DDPG is a typical RL algorithm for continual variable black-box optimization problem \cite{RL_overview}, enhancing the optimization performance through interactions with environments.
The scale of deep neural networks (DNNs) and hyper-parameters of DDPG are shown in Table. \ref{hyper_RL}. 

\item \textbf{LLM-generated non-dominated sorting genetic algorithm II (LNSGA):} 
In this method, the heuristic algorithm NSGA is recommended by LLMs, along with its programming code, heuristic rules, and hyper-parameters, based on the same task description prompt used in the LMWO and LHS frameworks.  
The key hyper-parameters of LNSGA are shown in Table. \ref{table:GA_parameters}.

\item \textbf{Simulated annealing genetic algorithm (SAGA):} This method is a classic heuristic algorithm combining simulated annealing and genetic algorithms \cite{paper_Lan2016_GASA_theory}.
The key hyper-parameters of SAGA are shown in Table. \ref{table:GA_parameters}.

\item \textbf{Random choice (RC):} All variables are randomly chosen within their respective solution space under constraints. 

\end{itemize}

\begin{table}
\begin{center}
\setlength{\abovecaptionskip}{1pt}
\setlength{\belowcaptionskip}{1pt}
\renewcommand{\arraystretch}{1.2}
\caption{Scale of DNNs and Hyper-parameters} 
\label{hyper_RL}
\centering
\begin{tabular}{|c|c|c|c|}
\hline
{\multirow{2}*{Actor DNNs}}  & Input layer & Hidden layers & Output layer \\
\cline{2-4}
{~} & Size of weight & 128 $\longrightarrow$ 64 & 3N   \\\hline
{\multirow{2}*{Critic DNNs}} & Input layer & Hidden layers & Output layer \\
\cline{2-4}
{~} & Size of weight + 3N & 128 $\longrightarrow$ 32 & 1      \\\hline
Replay buffer size & 5000  & Minibatch size & 32  \\\hline 
Soft update rates & 0.0001 &  Learning rates & 0.0001 \\\hline 
Discount factor &  0.99  &  & \\\hline 

\end{tabular}
\end{center}
\end{table}

\begin{table}[!t]
\begin{center}
\setlength{\abovecaptionskip}{1pt}
\setlength{\belowcaptionskip}{1pt}
\renewcommand{\arraystretch}{1.2}
\caption{LNSGA and SAGA Simulation Parameters} 
\label{table:GA_parameters}
\centering
\begin{tabular}{|c|c|c|c|}
\hline
Parameter & Value & Parameter & Value\\\hline
\text{Population size} & 50 & \text{Generations} & 1500 \\\hline 
\text{Crossover probability} & 0.7 & \text{Mutation probability} & 0.3 \\\hline 
Elite number & 2 & cooling factor & 0.98\\\hline 
high temperature & 100 & low temperature & 5 \\\hline 
\end{tabular}
\end{center}
\end{table}

Our simulations are executed using the Anaconda 3 software environment with Python 3.12.4 on a computer with a 2-core Intel i5-10505 3.20GHz CPU and 16 GB RAM.
For the daylight evaluation system, the Radiance 5.2.d and Daysim 4.0 software are adopted within Rhinoceros 8 SR7.
Additionally, the LLM component utilizes the Anthropic LLM via its commercial API \cite{anthropic_API}.

\begin{table}
\begin{center}
\setlength{\abovecaptionskip}{1pt}
\setlength{\belowcaptionskip}{1pt}
\renewcommand{\arraystretch}{1.2}
\caption{Simulation Parameters} 
\label{table:paremeters}
\centering
\begin{tabular}{|c|c|c|c|c|c|}
\hline
 Parameter & Value & Parameter & Value & Parameter & Value\\\hline
 $\lambda$ & 0.01 m & $\omega$ & 5 dB & $\textit{N}_{\rm t}$ & 32\\\hline
 $\textit{L}_{\rm r}$ & 20 m & $\textit{W}_{\rm r}$ & 10 m & $\textit{P}_{\rm t}$ & 1 W\\\hline
 $\textit{x}_{\rm b}$ & 20 m & $\textit{y}_{\rm b}$ & 30 m & $\textit{H}_{\rm b}$ & 30 m\\\hline 
 $L$ & 1.2 m & $W$ & 0.9 m & $\textit{H}_{{\rm t}}$ & 10 m \\\hline 
 $\textit{R}$ & 0.2 m & $\textit{G}$ & 8  & $\rho$ & $0.01  \ {\rm m}^{-2}$\\\hline
 $\textit{d}_x$ & 0.005 m & $\textit{d}_y$ & 0.005 m & $\textit{U}$ & 900\\\hline
 $\textit{F}_{\rm w}$ & $0.6$ & $\textit{F}_{\rm f}$ & 0.4 & $\textit{F}_{\rm c}$ & 0.7 \\\hline 
 $\eta$ & 5 & $\phi_{\rm d}$ & $220^{\circ}$ & $\theta_{\rm d}$ & $86^{\circ}$ \\\hline 
 $d_{\rm min}$ & 0.9 m & $\mathcal{T_{\rm min}}$ & 0.8 & $\mathcal{T}_{\rm i}$ & 5 \\\hline 
 
\end{tabular}
\end{center}
\end{table}

\subsection{Performance Analysis}
\begin{figure}[!t]
    \centering
    \includegraphics[width=3.10in]{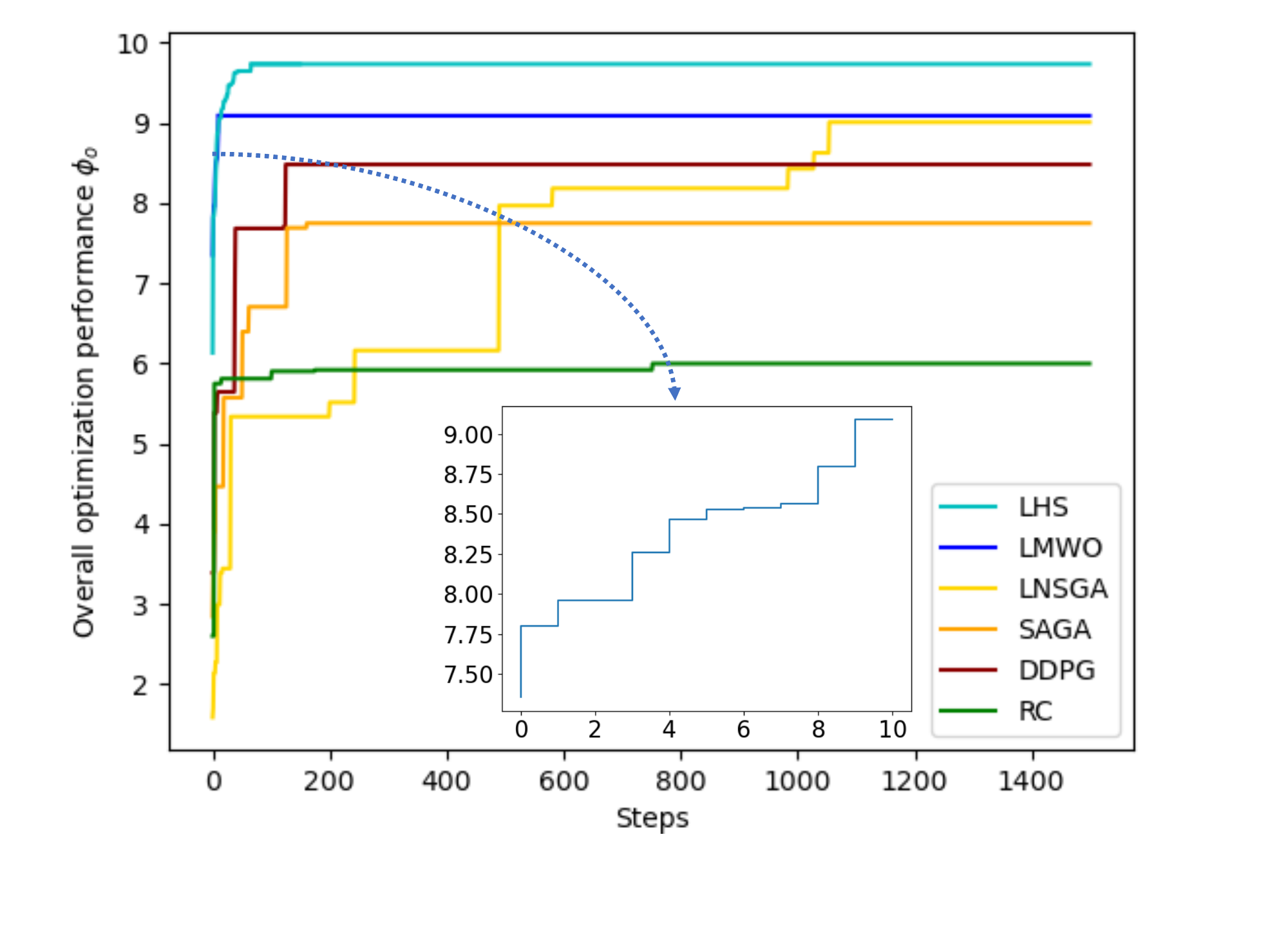}
    \caption{The average optimization performance of the LHS and LMWO framework compared with benchmark methods versus iteration steps with two windows.}
    \label{N_20_10_2win}
\end{figure}

\begin{figure}[!t]
    \centering
    \includegraphics[width=3.10in]{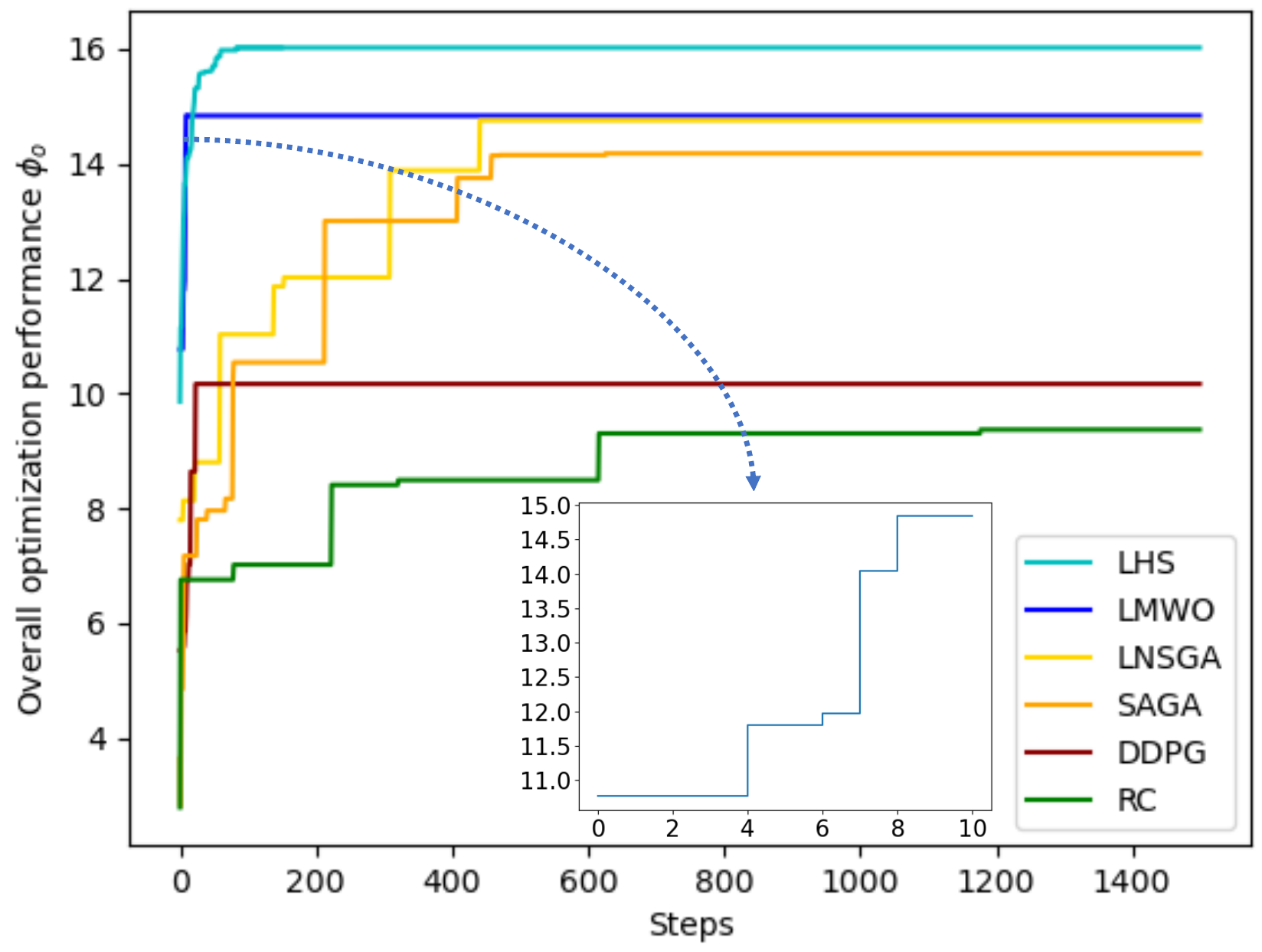}
    \caption{The average optimization performance of the LHS and LMWO framework compared with benchmark methods versus iteration steps with three windows.}
    \label{N_20_10_3win}
\end{figure}

\begin{figure}[!t]
    \centering
    \includegraphics[width=3.10in]{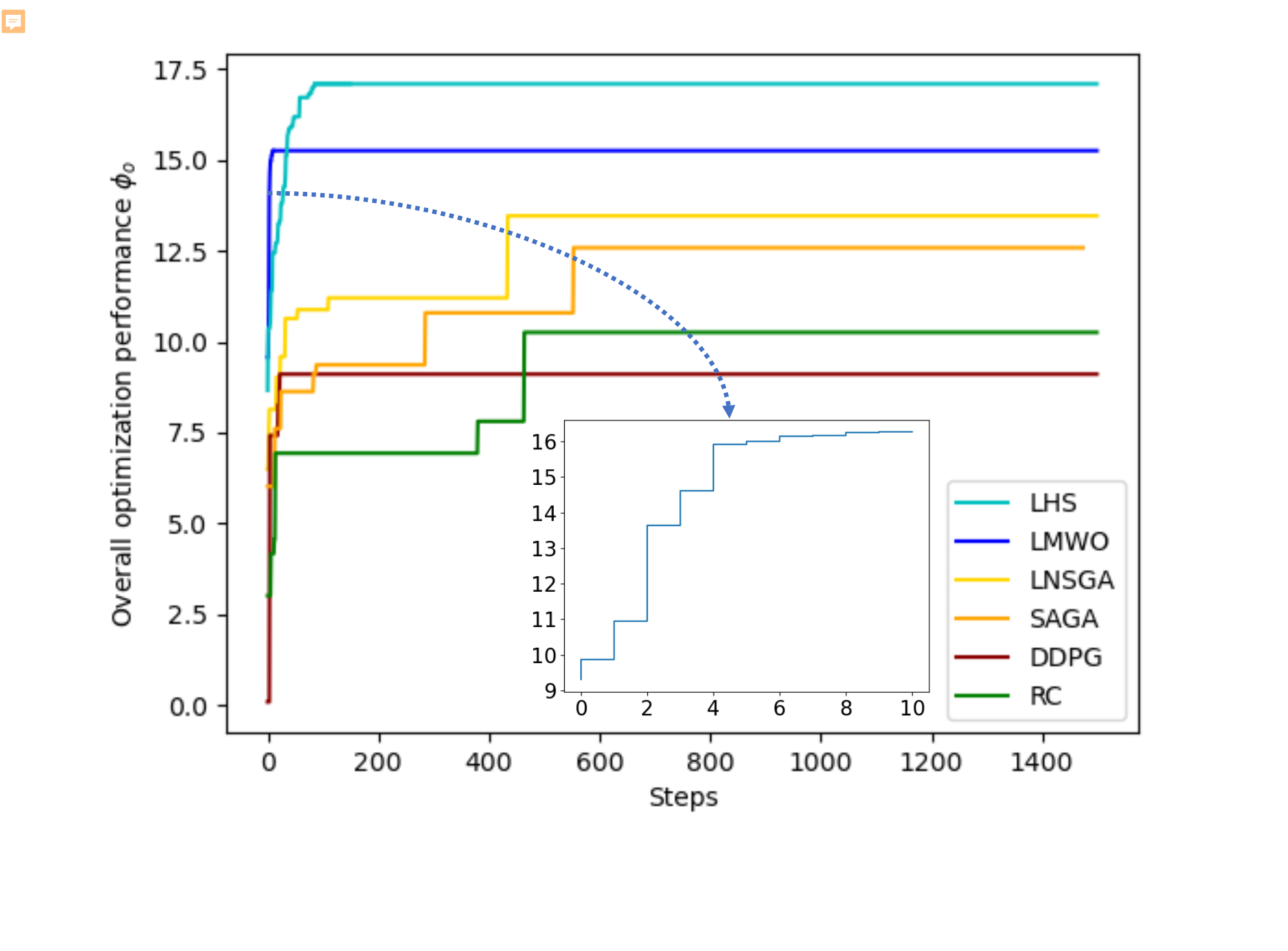}
    \caption{The average optimization performance of the LHS and LMWO framework compared with benchmark methods versus iteration steps with four windows.}
    \label{N_20_10_4win}
\end{figure}

In this subsection, we present the simulation results of the average optimization performance of the LMWO and LHS frameworks versus the iteration steps compared with benchmark methods.
Fig. \ref{N_20_10_2win} to Fig. \ref{N_20_10_4win} depict the average performance of the LMWO and LHS frameworks in a room with two, three, and four windows, respectively.
In all three figures, the LMWO and LHS frameworks consistently exhibit outstanding warm start and searching efficiency performance, significantly surpassing the other comparison methods. Meanwhile, they also achieve competitive or superior final performance compared to conventional heuristic algorithms.
The LMWO framework achieves faster convergence, whereas the LHS framework demonstrates superior final performance with more iterations.
For instance, in the four-window scenario, the LMWO framework starts from 8.5 and reaches a stable convergence in 10 iterations, achieving a final performance of 15.2. The LHS framework starts from 8.2 and reaches a stable convergence in 85 iterations, ultimately achieving a performance of 16.34.
In contrast, the performance of the SAGA and LNSGA methods is slightly inferior to the two aforementioned frameworks, but requires substantially more iterations to reach comparable results.
Additionally, the DDPG method only achieves performance comparable to the heuristic baseline algorithm in the two-window scenario, which degrades as the number of windows increases. 
Finally, the RC method exhibits the poorest performance across all the scenarios.

All three scenarios demonstrate that the LMWO framework can effectively harness the reasoning capabilities of LLMs and continuously optimize the output with outstanding wireless network planning capacity and the architecture daylight planning capacity, coupled with a rapid exploration speed. 
Upon integration with heuristic strategies, the LHS framework benefits from LLM-generated high-quality initialization, enabling the heuristic algorithms to achieve approximately a 90\% improvement in convergence speed and a 15\% increase in final performance.
Moreover, the results clearly illustrate a substantial optimization potential for window design to improve the overall indoor QoE compared with the UWD strategy.

To our best knowledge, three main factors contribute to the powerful optimization performance of the LMWO and LHS frameworks.
Firstly, the pre-trained wireless and architecture knowledge embedded in the LLM enables the LMWO and LHS frameworks to understand their tasks and propose robust initial solutions without additional training.
Secondly, the LMWO and LHS frameworks efficiently extract relevant information from our designed prompts, empowered by LLMs' semantic understanding and reasoning capacity, leading to progressively improved solutions.
Finally, the human-like continual learning ability of LLMs enables them to incorporate external multi-modal feedback from the evaluation system, thereby iteratively refining the optimization outcomes.

\subsection{The Influence of Windows Number}
\begin{figure}[!t]
    \centering
    \includegraphics[width=3.10in]{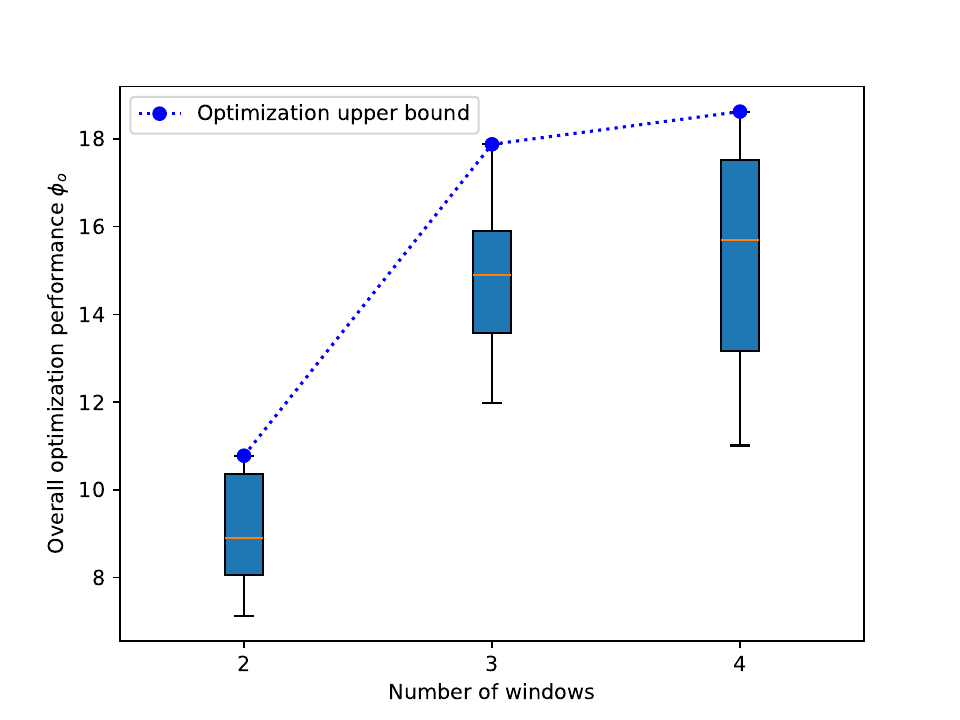}
    \caption{The reliability and the best optimization performance of the LMWO framework versus the number of windows.}
    \label{N_20_10}
\end{figure}

Firstly, we investigate the influence of the number of windows on the performance and the reliability of the LMWO and LHS frameworks by considering two, three, and four windows. 
Fig. \ref{N_20_10} illustrates the influence of the number of windows on the upper bound of optimization performance, the median value, and the reliability of the LMWO framework.  
As the number of windows increases from 2 to 4, the upper bound of optimization performance rises from 10.8 to around 18.6, and the average performance rises from 9.1 to 15.3. 
However, the performance variability also becomes significantly larger.
This observation indicates that while a greater number of windows enhances the potential for higher optimization performance, it also reduces reliability. 
This is because increasing the number of windows expands the solution space, thereby making the exploration process more challenging and more susceptible to local suboptimal solutions.

\begin{figure}[!t]
    \centering
    \includegraphics[width=3.10in]{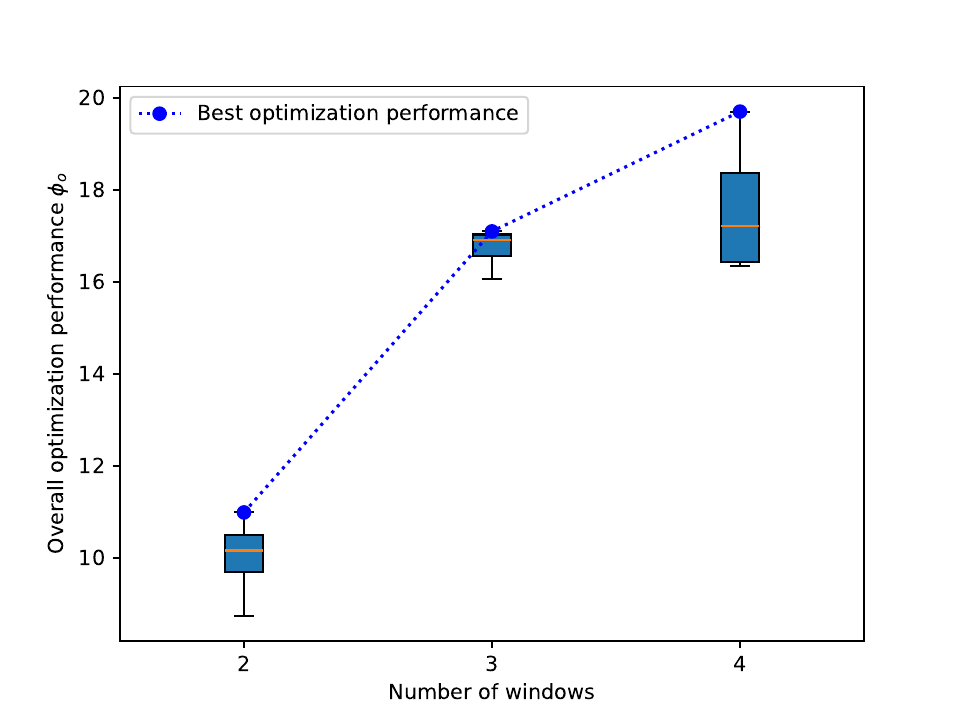}
    \caption{The reliability and the best optimization performance of the LHS framework versus the number of windows.}
    \label{LHS_N_20_10}
\end{figure}

Next, Fig. \ref{LHS_N_20_10} illustrates the influence of the number of windows on the upper bound of optimization performance, the median value, and the reliability of the LHS framework.
As the number of windows increases from 2 to 4, the upper bound of optimization performance rises from 10.9 to 19.7, and the average performance rises from 9.7 to 17.1. 
Compared with the LMWO framework, the LHS framework demonstrates greater reliability in performance after the integration of heuristic algorithms.
This improvement is attributed to the fact that heuristic algorithms introduce a coherent and structured search strategy, which effectively mitigates the variability and instability commonly associated with LLMs across diverse scenarios.
Consequently, the LHS framework exhibits enhanced capability in avoiding suboptimal solutions and conducting a more exhaustive search at a cost of a slightly increased computational complexity compared with the LMWO framework.

\subsection{The Influence of RIS Units Number}
\begin{figure}[!t]
    \centering
    \includegraphics[width=3.4in]{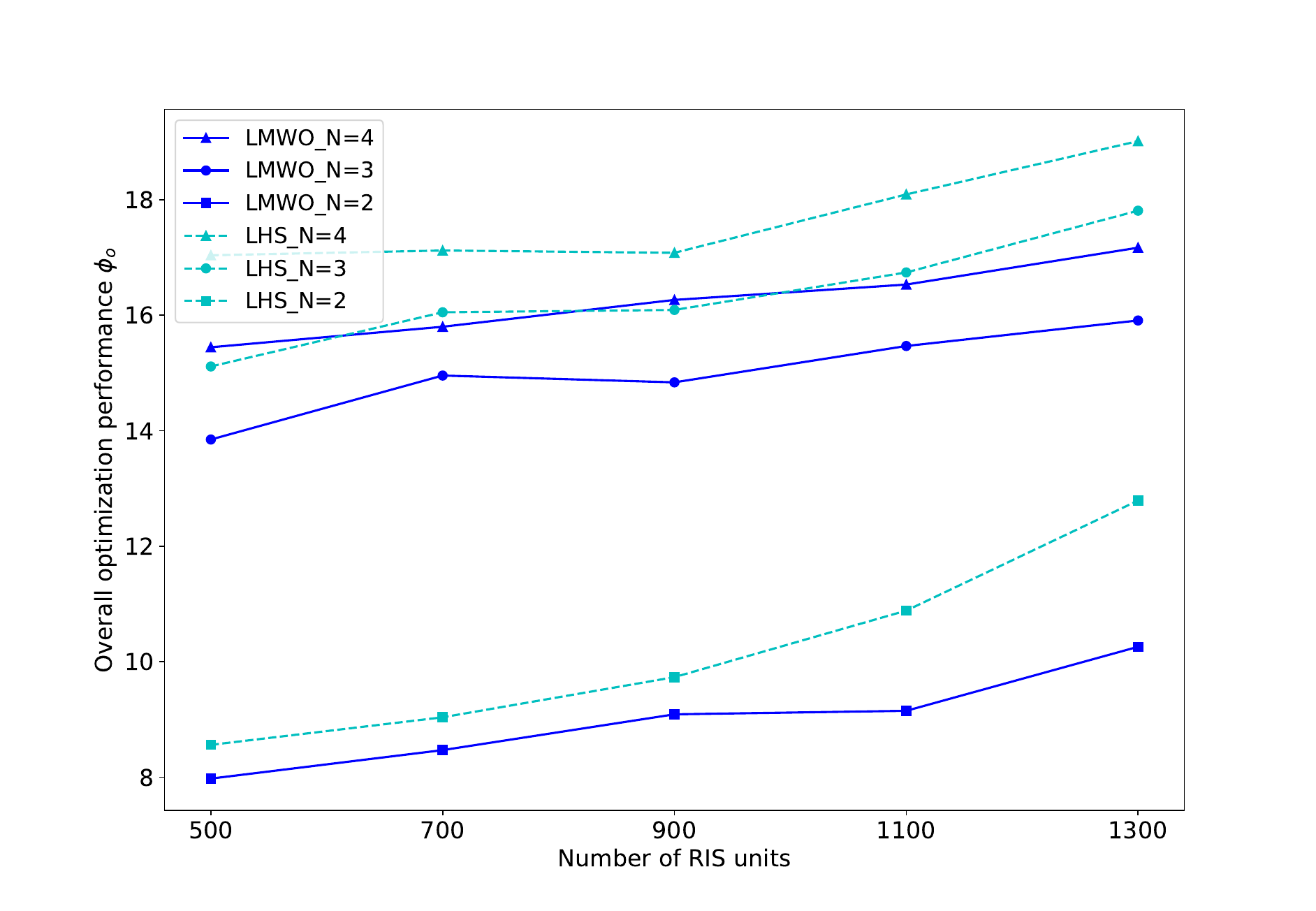}
    \caption{The average performance improvement achieved by the LMWO and LHS frameworks versus the total units of RISs.}
    \label{RIS_unit_1}
\end{figure}

We investigate the influence of the number of RIS units on the optimization performance of the LMWO and LHS frameworks. 
As the number of RIS units increases, the number of RIS units allocated per window also rises. 
Consequently, the orientations of RISs play an increasingly critical role in the optimization process, and wireless performance improvement becomes more important.
Specifically, Fig. \ref{RIS_unit_1} demonstrates the influence of the number of RIS units on the average optimization performance of the LMWO and LHS frameworks, considering 500, 700, 900, 1100, and 1300 units with two, three, and four windows.
In all these scenarios, the average optimization performance exhibits a slightly increasing trend as the number of RIS units increases. However, a notable improvement in optimization performance is observed only in scenarios with two windows.
It demonstrates that increasing the number of RIS units primarily enhances wireless network performance, which has a greater influence on the overall optimization outcome than improvements in daylight performance.

Moreover, in scenarios with three and four windows, the performance of the LMWO and LHS frameworks is comparable and consistently outperforms that of the two-window scenario in all RIS unit scenarios. 
This observation demonstrates that both the LMWO and LHS frameworks can achieve strong optimization performance regardless of the number of RIS units.
It also suggests that a greater number of windows provides increased deployment flexibility, allowing more effective utilization of RIS units and leading to improved overall optimization performance.
Additionally, the LHS framework consistently outperforms the LMWO framework, indicating that the integration of heuristic strategies allows the LHS framework to achieve superior solutions across varying RIS unit configurations.

\subsection{The Influence of Daylight Factor}
\begin{figure}[!t]
    \centering
    \includegraphics[width=3.25in]{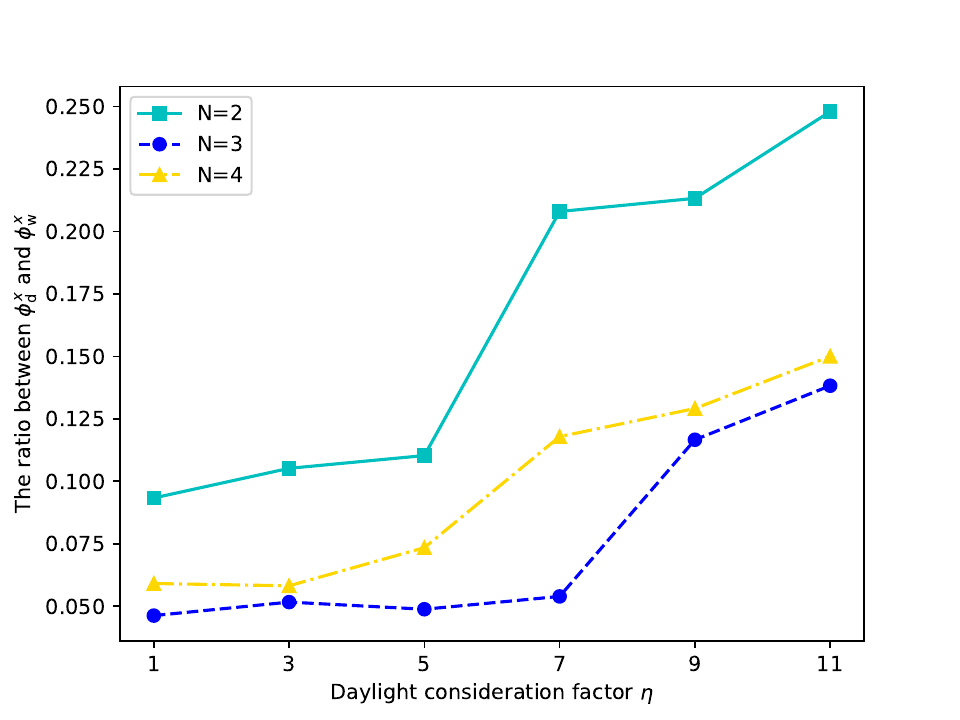}
    \caption{The ratio between the daylight performance improvement and wireless performance improvement achieved by the LMWO framework versus the daylight factor.}
    \label{eta_20_10}
\end{figure}

Furthermore, we investigate the influence of the daylight factor $\eta$ on the daylight optimization tendency of the LMWO and LHS frameworks, quantified by the ratio of daylight performance improvement $\phi_{\rm d}^{x}$ to wireless performance improvement $\phi_{\rm w}^{x}$. The daylight factor varies from one to eleven in increments of two. 
Specifically, Fig. \ref{eta_20_10} demonstrates the influence of the daylight factor on the daylight optimization tendency of the LMWO framework when there are two, three, and four windows. 
We notice that as the daylight factor $\eta$ increases, the daylight optimization tendency significantly increases across all scenarios. In the two-window case, the daylight optimization tendency remains around 0.1 from daylight factors 0 to 3, followed by a dramatic increase from daylight factor 3 to 11. 
Moreover, with three and four windows, the daylight optimization tendency gradually increases as the daylight factor increases.
These results demonstrate that the LMWO method effectively considers the daylight factor, exhibiting a clear daylight optimization tendency and semantic understanding capacity. 
Moreover, the performance of the LMWO framework appears to be more sensitive to variations in the daylight factor when fewer windows are present, exhibiting a steep increase in daylight optimization tendency in the two-window scenario. Conversely, in the four-window configuration, the performance trend is smoother.
This is because fewer windows offer more flexibility in adjusting their positions on the wall, enabling the LMWO framework to more easily identify optimal solutions for enhancing daylight performance, but potentially at the expense of wireless performance.

\begin{figure}[!t]
    \centering
    \includegraphics[width=3.25in]{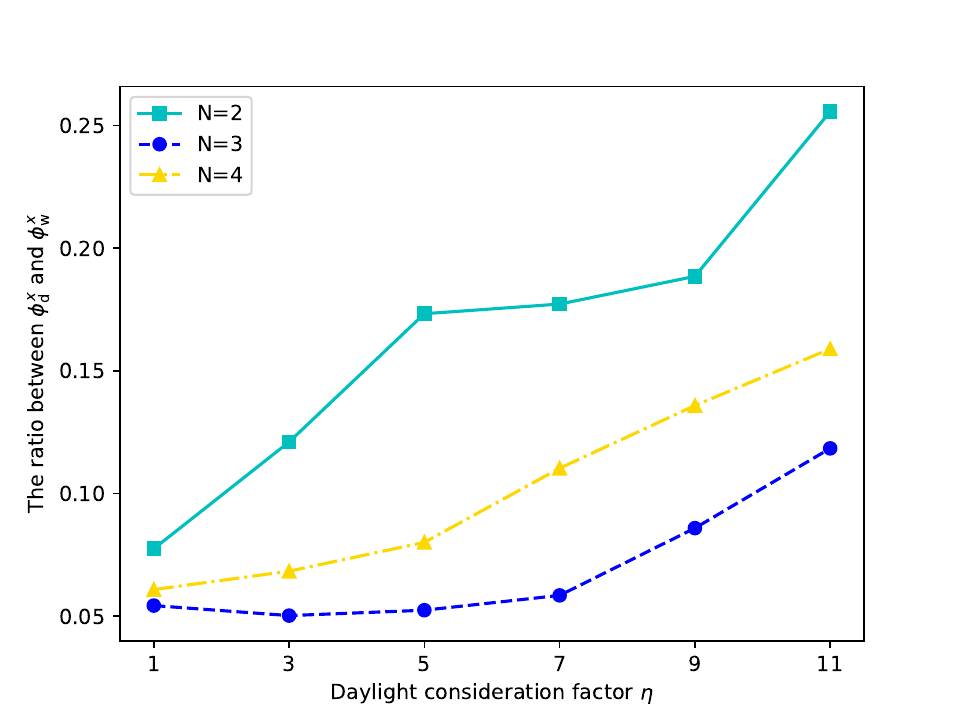}
    \caption{The ratio between the daylight performance improvement and wireless performance improvement achieved by the LHS framework versus the daylight factor.}
    \label{eta_20_10_LHS}
\end{figure}

Similar phenomena are observed in Fig. \ref{eta_20_10_LHS} for the LHS framework. As the daylight factor varies from 1 to 11, the ratio of daylight performance improvement $\phi_{\rm d}^{x}$ to wireless performance improvement $\phi_{\rm w}^{x}$ keeps increasing in all window quantity scenarios.
This trend demonstrates that the LHS framework can also consider optimization preference in initialization and optimization steps guided by LLMs, even though the search process is conducted by the proposed heuristic algorithms.  
Furthermore, the LHS framework demonstrates a similarly sharp performance improvement in daylight optimization tendency in the two-window case, while the scenarios with three and four windows exhibit more moderate and gradual increases. This supports that fewer windows offer greater flexibility in adjusting their positions on the wall to enhance daylight performance, but significantly reduce the wireless performance.
Therefore, it can be concluded that more windows facilitate a more effective balance between daylight illumination and wireless communication performance.

\subsection{Time Complexity}

\begin{table}
\begin{center}
\setlength{\abovecaptionskip}{1pt}
\setlength{\belowcaptionskip}{1pt}
\renewcommand{\arraystretch}{1.2}
\caption{Average Operation Time Costs} 
\label{table:time}
\centering
\begin{tabular}{|c|c|c|c|}
\hline
Methods & $N$ = 2 & $N$ = 3 & $N$ = 4\\\hline
LMWO & 73.2 s & 89.7 s & 92.3 s\\\hline
LHS &  536.6 s & 753.5 s & 784.6 s  \\\hline
LNSGA & 21916.5 s & 9399.9 s & 9261.2 s \\\hline
SAGA & 3365.4 s & 13274.2 s & 11794.7 s \\\hline
DDPG & 229.5 s & 49.8 s & 49.1 s \\\hline
RC & 1371.1 s & 1278.1 s & 990.7 s \\\hline

\hline
\end{tabular}
\end{center}
\end{table}

Eventually, we examine the time complexity of each method, including the online operation process and the environmental interaction process, as shown in Table \ref{table:time}. 
We observe that the time complexity of the LHS and LMWO frameworks increases as the number of windows increases.
This is because these two frameworks spend more time searching for solutions in a bigger solution space.
The time complexity of the LNSGA, SAGA, DDPG, and RC methods decreases as the number of windows increases because the expansion of the solution space increases the search difficulty, leading to earlier convergence.
Although the LMWO and LHS frameworks incur a relatively higher online time cost compared to other methods due to the internet speed and the sequence generation speed of LLMs, their superior reasoning efficiency still provides significant time advantages. 
This is because the LHS and LMWO methods only require fewer iterations to achieve satisfactory stable results, benefiting from the pre-trained knowledge and fast learning capacity from feedback, instead of hundreds of iterations in other methods. 

Additionally, we need to emphasize that the LMWO and LHS frameworks involve extra API cost for cloud-based commercial LLMs \cite{anthropic_API}, which is determined by the input and output size and the LLM model applied. Thus, it is important to balance the optimization effectiveness, efficiency, and cost by adjusting the prompt in practical use \cite{LLM_cost}.

\section{Conclusion}    
This paper has addressed a joint optimization of indoor wireless and daylight performance in RIS-aided O2I scenarios. 
We have highlighted the significance of the window-deployment strategy of T-RIS and formulated an optimization problem that simultaneously considers the wireless and daylight performance.  
Moreover, a novel optimization framework named LMWO has been proposed to tackle this problem based on the multi-modal LLM technology using prompt engineering technology.
Another LHS framework is also proposed, which integrates the comprehensive search strategies of heuristic algorithms to enhance optimization performance and results stability.
Our numerical results have revealed that our proposed LMWO framework significantly improves the overall performance of both wireless and daylight, illustrating the critical role of window placement compared to the uniform distribution strategy.
Furthermore, the LMWO framework has shown substantial optimization performance in warm start, exploration speed, and final performance, compared with the SAGA heuristic optimization algorithm.
In addition, we have analyzed the impact of the number of windows, number of RIS units, room size, and daylight factor on the performance of the LMWO framework, showcasing its advanced optimization capabilities and the strong generalization ability of our proposed framework. 
These findings have underscored the significance of window positioning and the potential of LLMs as optimizers in RIS-aided O2I wireless communications scenarios.

\bibliography{Reference}
\bibliographystyle{IEEEtran}  


\end{document}